\pdfoutput=1
\documentclass[11pt]{article}
\usepackage[pdftex]{graphicx,color} 
\usepackage{jheppub}
\usepackage{amsmath}
\usepackage{amssymb}
\usepackage{comment}
\usepackage{multirow}
\usepackage{mathtools}
\usepackage{dsdshorthand}
\usepackage{shuffle}
\newcommand{\bea}{\begin{equation}\begin{aligned}}
\newcommand{\eea}[1]{\label{#1}\end{aligned}\end{equation}}
\newcommand{\beq}{\begin{equation}}
\newcommand{\eeq}{\end{equation}}
\newcommand   \xb  {\bar{x}}
\newcommand   \zb  {\bar{z}}

\title{Single-valuedness of the AdS Veneziano amplitude}
\author[a]{Luis F. Alday}
\author[b]{and Tobias Hansen}

\affiliation[a]{Mathematical Institute, University of Oxford,
Woodstock Road, Oxford, OX2 6GG, UK}
\affiliation[b]{Department of Mathematical Sciences, Durham University,
Stockton Road, Durham, DH1 3LE, UK}
\emailAdd{luis.alday@maths.ox.ac.uk}
\emailAdd{tobias.p.hansen@durham.ac.uk}

\abstract{
We consider the Veneziano amplitude for the scattering of gluons in type IIB string theory on $AdS_5 \times S^5/\mathbb{Z}_2$ in the presence of D7 branes. On general grounds curvature corrections around flat space can be thought of as arising from the extra insertion of soft gravitons. This naturally leads to an open string world-sheet representation with the extra insertion of single-valued functions evaluated on the real line. We check that the recently obtained first curvature correction is of this form and use this new constraint to compute the second curvature correction of the AdS Veneziano amplitude.
}

\begin{document}
\maketitle

\section{Introduction}

In this paper we consider the scattering of massless open strings on $AdS_5 \times S^3$ in a small curvature expansion. In the simplest holographic setup one considers a small number of D7 branes as probes to $AdS_5 \times S^5$.\footnote{See \cite{Alday:2024yax,Aharony:1998xz,Ennes:2000fu,Karch:2002sh,Alday:2021odx,Behan:2023fqq} for further details on the setup.} The world-volume of the D7-branes wraps around $AdS_5$ and an equatorial $S^3$ inside $S^5$. The open strings whose scattering we consider are attached to these D7 branes. In flat space and for four points the result is simply the Veneziano amplitude. The world-sheet has the topology of a disk which can be mapped to the upper half plane, with the open string vertex operators situated on the real line. Choosing the locations at $(0,x,1,\infty)$ one obtains the familiar expression 
\begin{equation}
\int_0^1 dx \, \langle V_o(0) V_o(x) V_o(1) V_o(\infty) \rangle \sim \int_0^1 dx \, x^{-S-1}(1-x)^{-T-1}\,.
\end{equation}
Curvature corrections are equivalent to the insertion of extra gravitons with soft momenta, as one can see by considering a toy non-linear $\sigma$-model with AdS expanded around flat space as target space \cite{Alday:2023jdk,Alday:2023mvu}. Mixed scattering of open and closed massless strings in flat space was thoroughly considered in \cite{Stieberger:2009hq,Stieberger:2015vya}. For the insertion of an extra graviton at a location $z$ one obtains integrals of the form  
\begin{align}
&\int_0^1 dx \,\int_{H^+} d^2 z \,  \langle V_o(0) V_o(x) V_o(1) V_o(\infty) V_{c}(z,\bar z) \rangle \\
&\sim \int_0^1 dx \, x^{-S-1}(1-x)^{-T-1}  \int_{H^+} d^2 z \frac{|z|^{2 \alpha' p_1 q}|1-z|^{2 \alpha' p_3 q}(x-z)^{\alpha' p_2 q}(x-\bar z)^{\alpha' p_2 q}(z-\bar z)^{\alpha'q^2_{\parallel}}}{|z|^2|1-z|^2 } \,,
\nonumber
\end{align}
where $q$ is the momentum of the soft graviton, with $q_\parallel$ its longitudinal component along the brane.  The integral over the location of the graviton vertex operator is an integral over the upper half plane. As explained in \cite{Stieberger:2009hq,Stieberger:2015vya} however, the integration can be extended to the full complex plane in the case of symmetric closed string states, such as the graviton. Furthermore, in the context of this paper, the full momentum of the soft graviton lies on the D7 brane, since we are computing $AdS_5 \times S^3$ curvature corrections, so that $q^2_{\parallel}=0$. In conclusion, the insertion of an extra graviton leads to the extra insertion of ${\cal J}(x,x)$, the function ${\cal J}(x,\bar x)$ evaluated on the real line $x=\bar x$, with 
\begin{equation}
{\cal J}(x,\bar x) = \int_\mathbb{C} d^2z \frac{|z|^{2a}|1-z|^{2b}|z-x|^{2c}}{|z|^2|1-z|^2}\,, \qquad (a,b,c)=\alpha'(p_1 q,p_3 q,p_2 q)\,.
\label{J}
\end{equation}
This is the same type of insertions considered in \cite{Alday:2023jdk}, where it was argued that in the small $q$ expansion this produces single-valued multiple polylogarithms (SVMPLs). The computation of \cite{Alday:2023jdk} is reproduced and extended in the appendix. 

The following picture then emerges. AdS curvature corrections to the  Veneziano amplitude lead to the extra insertion of single-valued multiple polylogarithms, evaluated on the real line $\left. {\cal L}_w(x) \right|_{\xb=x}$. This is a particular class of multiple polylogarithms (MPLs). To summarise, AdS curvature corrections correspond to the insertion of extra (soft) gravitons, and the scattering of gravitons is governed by single-valued insertions.\footnote{Note that single-valuedness is only a property of the integrand. For the amplitude itself,  single-valuedness is lost because of the one-dimensional $x$ integral. In particular, the low energy expansion coefficients of the amplitude contain non-single-valued multiple zeta values (MZVs). This is the usual notion of the absence of single-valuedness in the Veneziano amplitude, already apparent in flat space.}

This insight will be a powerful ingredient in the construction of the AdS Veneziano amplitude, as recently initiated in \cite{Alday:2024yax} following up on similar work on the AdS Virasoro-Shapiro amplitude \cite{Alday:2022uxp,Alday:2022xwz,Alday:2023jdk,Alday:2023mvu}. The AdS amplitude $A(S,T)$ is defined in terms of the holographic correlator of four flavour currents in the dual $\mathcal{N}=2$ superconformal field theory and satisfies a dispersion relation, expressing it in terms of OPE data of massive string operators. In addition, it admits a representation as an open string world-sheet integral in an expansion around flat space
\beq
A(S,T) = \frac{1}{S+T} \int\limits_0^1 d x \, x^{-S-1} (1-x)^{-T-1} 
\sum\limits_{k=0}^{\infty} \left(\frac{\alpha'}{R^2} \right)^k G^{(k)} (S,T,x)  \,,
\eeq
where $R$ is the AdS radius and the leading term is the flat space Veneziano amplitude
\beq
G^{(0)} (S,T,x) = 1\,.
\eeq
The first curvature correction $G^{(1)} (S,T,x)$ was fully determined in \cite{Alday:2024yax} by using an ansatz for this function in terms of multi-valued MPLs, however such an ansatz did not lead to a unique solution for the next correction $G^{(2)} (S,T,x)$.
The first correction, obtained without assuming single-valuedness, can actually be expressed in terms of SVMPLs on the real line\footnote{We thank Pietro Ferrero for pointing this out to us.}
\begin{align}
\label{G1_SVMPLs}
G^{(1)} (S,T,x) ={}&
\frac{S^2+T^2}{4} \left(\mathcal{L}^+_{000}(x)-\mathcal{L}^+_{001}(x) \right)
-\frac{(S+T)^2}{4} \left(\mathcal{L}^+_{010}(x) - 4 \zeta (3)\right)\\
&-\frac{3 S^2+8 S T+3 T^2}{4(S+T)} \mathcal{L}^+_{00}(x)
+\frac{5 S^2+12 S T+5 T^2}{4(S+T)} \mathcal{L}^+_{01}(x)
+\frac34 \mathcal{L}^+_0(x) 
+\frac{3}{S+T}\nonumber\\
&+\frac{S-T}{4} \left((S+T) \left(\mathcal{L}^-_{000}(x)-\mathcal{L}^-_{001}(x)-\mathcal{L}^-_{010}(x)\right)-3
   \mathcal{L}^-_{00}(x)-\frac{5 \mathcal{L}^-_0(x)}{S+T}\right)\,,\nonumber
\end{align}
where $\bar x=x$ and
\beq
\mathcal{L}^\pm_{w}(x) \equiv \mathcal{L}_{w}(x) \pm \mathcal{L}_{w}(1-x)\,. 
\eeq
This provides further evidence for the claim that the insertions are single-valued. As it turns out, the assumption of single-valuedness is enough to fully fix the second curvature correction $G^{(2)} (S,T,x)$.

The remainder of the paper is organised as follows.
In Section \ref{sec:setup} we review the holographic correlator and the dispersion relation. 
Section \ref{sec:world-sheet_correlator} discusses the space of SVMPLs on the real line and the solution for $G^{(2)} (S,T,x)$.
In Section \ref{sec:checks} we perform consistency checks by considering the high energy limit, the low energy expansion and the OPE data of the exchanged massive string operators.
We find that the high energy limit is consistent with exponentiation, as predicted in \cite{Alday:2023pzu} in the context of closed strings. In the low energy expansion we fully fix the unprotected $D^6 F^4$ term in the holographic correlator.
The OPE data can be compared to a classical open string solution in AdS and we find agreement.
We conclude in Section \ref{sec:conclusions}.
Finally, the small $q$ expansion of \eqref{J} is done in Appendix \ref{app:J}.

\section{Setup}
\label{sec:setup}

\subsection{Correlator}

The holographic dual of the string theory under consideration is a 4d $\mathcal{N} = 2$ superconformal field theory with $R$-symmetry group $SU(2)_R \times U(1)_r$ and flavour symmetry $SU(2)_L \times G_F$, where $G_F$ can be $SO(8)$, $U(4)$ or $SO(4)\times SO(4)$.\footnote{These cases were discussed in \cite{Ennes:2000fu}.}
We consider the correlator of four moment map operators $\mathcal{O}^I_\alpha(x)$, which is the superconformal primary of the flavour supermultiplet, a Lorentz scalar with dimension $\Delta=2$ in the singlet of $SU(2)_L$, the adjoint of $SU(2)_R$ (with index $\alpha$) and also in the adjoint of $G_F$ (with index $I$). 
In particular we work with the reduced Mellin amplitude $M^{I_1 I_2 I_3 I_4}(s, t)$ for this correlator.\footnote{See \cite{Alday:2024yax} for the relation to the correlator.}
We only consider the leading contribution of order $1/N$ in the large $N$ expansion\footnote{We will drop the overall factor $1/N$ from formulas.}, which corresponds to open string scattering where the world-sheet has the topology of a disc with four insertions at the boundary. The colour structures for this configuration are single traces of the generators $T^I$ of $G_F$ and the amplitude takes the form \cite{Alday:2021odx,Behan:2023fqq}
\bea
M^{I_1 I_2 I_3 I_4}(s, t) ={}& 
{\rm Tr} \left(T^{I_1}T^{I_2}T^{I_3}T^{I_4} \right)
  M(s,t)
+{\rm Tr} \left(T^{I_1}T^{I_4}T^{I_2}T^{I_3} \right)
M(t,u) \\
&+ {\rm Tr} \left(T^{I_1}T^{I_3}T^{I_4}T^{I_2} \right)
M(u,s)
\,,
\eea{trace_basis}
where $M(s,t)$ is called the colour-ordered Mellin amplitude and can have only poles in the $s$- and $t$-channels, the only ones consistent with the colour ordering $(1234)$. Crossing symmetry implies that
\beq
M(s,t) =  M(t,s)\,.
\eeq
Apart from the Mellin amplitude we will also study its Borel transform
\beq \label{borel}
A(S,T) \equiv \frac{\lambda}{8} \int_{\kappa-i\infty}^{\kappa+ i \infty} \frac{d\alpha}{2 \pi i} \, e^\alpha \alpha^{-4} 
M \left( \frac{\sqrt{\lambda} T}{2\alpha}, \frac{\sqrt{\lambda} U}{2\alpha} \right)  \,,
\eeq
where the t'Hooft coupling $\lambda$ is related to the AdS radius $R$ and the Regge slope $\alpha'$ via the AdS/CFT dictionary
\beq
\sqrt{\lambda} = \frac{R^2}{\alpha'} + O\left(\frac{1}{N}\right) \,.
\label{dictionary}
\eeq
We call $A(S,T)$ the AdS amplitude.
When considering its large $\lambda$ expansion
\beq
A(S,T)= \sum_{k=0}^\infty \frac{1}{\lambda^\frac{k}{2}} A^{(k)}(S,T) \,,
\label{fsl}
\eeq
the leading term is the flat space Veneziano amplitude
\bea
A^{(0)}(S,T) ={}&
= -\frac{\Gamma(-S) \Gamma(-T)}{\Gamma(1-S-T)}\,.
\eea{flat_space_amplitude}
Because of this, \eqref{borel} is also known as the flat space limit \cite{Penedones:2010ue,Fitzpatrick:2011hu}.

\subsection{Dispersion relation}

As shown in \cite{Alday:2024yax}, the Mellin amplitude obeys a dispersion relation
\beq
M(s,t) 
=\sum_{\tau,\ell,m}\frac{ C_{\tau, \ell}^{2} \mathcal{Q}_{\ell, m}^{\tau +2, d=4}(t-2)}{s-\tau - 2m+2}\,,
\label{dr}
\eeq
where $\mathcal{Q}_{\ell, m}^{\tau +2, d=4}(t-2)$ is a Mack polynomial, defined as in \cite{Alday:2024yax}, and
the sum runs over massive string operators in the $s$-channel, whose dimensions are determined at large $\lambda$ by the mass levels $\alpha' m^2 = \delta=1,2,\ldots$ of the flat space string spectrum
\beq
\Delta = m R \left(1+ O\left( \lambda^{-\frac12} \right) \right) = \sqrt{\delta}\lambda^{\frac14} + O\left( \lambda^{-\frac14} \right)\,.
\eeq
We will expand the OPE data in large $\lambda$ as 
\bea
\tau(\delta,\ell) &= \sqrt{\delta} \lambda^{\frac{1}{4}} + \tau_1(\delta,\ell)  +  \tau_2(\delta,\ell) \lambda^{-\frac{1}{4}} + \ldots \,,\\
C^2_{\tau,\ell} &=  \frac{2^{-2\tau(\delta,\ell)-2\ell-9} \pi^3 \tau(\delta,\ell)^4}{(\ell+1) \sin(\frac{\pi}{2} \tau(\delta,\ell))^2}  f(\delta,\ell)\,,  \\
 f(\delta,\ell) &= f_0(\delta,\ell) +  f_1(\delta,\ell) \lambda^{-\frac{1}{4}} + f_2(\delta,\ell) \lambda^{-\frac{1}{2}} + \ldots \,,
\eea{ope_data_expansion}
where the flat space spectrum implies
\beq
 f(\delta,\ell) = 0\,, \quad \delta - \ell \text{ even}\,.
\label{flat_spectrum}
\eeq
By performing a large $\lambda$ expansion, the sum over $m$ in \eqref{dr} as well as the Borel transform \eqref{fsl} (see \cite{Alday:2022xwz,Fardelli:2023fyq}), we can obtain the following version of the dispersion relation for the AdS amplitude close to a pole $S \sim \delta$
\beq
A^{(k)}(S,T) = \sum\limits_{i=1}^{3k+1} \frac{R^{(k)}_i(T,\delta)}{(S-\delta)^i} + O((S-\delta)^0)\,.
\label{A_dr_poles}
\eeq
The numerators can be explicitly computed and depend linearly on the OPE data: $R^{(0)}_i(T,\delta)$ depends on the leading OPE coefficients $\langle f_0 \rangle_{\delta, \ell}$ and $R^{(1)}_i(T,\delta)$ depends on 
$\langle f_2 \rangle_{\delta, \ell}$ and $\langle f_0 \tau_2 \rangle_{\delta, \ell}$.
This data has been computed in \cite{Alday:2024yax} from $A^{(0)}(S,T)$ and $A^{(1)}(S,T)$.
In the present paper we compute $A^{(2)}(S,T)$ and therefore $\langle f_4 \rangle_{\delta, \ell}$, $\langle f_0 \tau_4 + f_2 \tau_2  \rangle_{\delta, \ell}$ and $\langle f_0 \tau_2^2 \rangle_{\delta, \ell}$.
The angle brackets indicate a sum over all operators that are degenerate in the flat space limit.

Corrections to the OPE data at odd powers of $\lambda^{1/4}$ are fixed by similar relations to \eqref{A_dr_poles} where the left hand side vanishes, because $A(S,T)$ can be expanded in integer powers of $\alpha'/R^2$.
This leads to
\begin{align}
\tau_1(\delta, \ell)  = - \ell\,, \qquad \langle f_1 \rangle_{\delta, \ell} = \langle f_0 \rangle_{\delta, \ell} \frac{4 \ell-\frac12}{\sqrt{\delta }}\,,
\label{tau1f1}
\end{align}
and at order $1/\lambda^{3/4}$ to
\bea
\tau_3(\delta, \ell)  ={}& 0\,,\quad
\langle f_1 \tau_2 \rangle_{\delta, \ell} = \langle f_0 \tau_2 \rangle_{\delta, \ell} \frac{4 \ell-\frac12}{\sqrt{\delta }}\,,\\
\langle f_3 \rangle_{\delta, \ell} ={}& 
\frac{4   (8 \ell-1) \left( \delta \langle f_2 \rangle_{\delta, \ell}-\sqrt{\delta }  \langle f_0 \tau_2 \rangle_{\delta, \ell}\right)- \left(160 \ell^3-60 \ell^2+16 \ell+3\right)\langle f_0 \rangle_{\delta, \ell}}{8 \delta ^{3/2}}\,.
\eea{OPE_data_3/2}

\section{World-sheet correlator}
\label{sec:world-sheet_correlator}

Our strategy to fix a given curvature correction $A^{(k)}(S,T)$ is to make the assumption that it has a representation as the world-sheet integral
\beq
A^{(k)}(S,T) = \frac{1}{S+T} \int\limits_0^1 dx \, x^{-S-1} (1-x)^{-T-1}
G^{(k)} (S,T,x)\,.
\label{worldsheet_integral}
\eeq
We then make an ansatz for $G^{(k)} (S,T,x)$ and demand consistency with \eqref{A_dr_poles} to fix the coefficient of the ansatz.

\subsection{Basis of SVMPLs on the real line}

In order to write a maximally restrictive ansatz for $G^{(k)} (S,T,x)$ we have to find a basis of SVMPLs\footnote{See \cite{Brown:2004ugm,Dixon:2012yy,Alday:2023jdk} for details on SVMPLs. We found the programs \cite{Duhr:2019tlz,Panzer:2014caa,HyperlogProcedures} useful for practical applications.} restricted to the real line $x = \xb$.
In \cite{Alday:2024yax} $G^{(1)} (S,T,x)$ was computed without assuming single-valuedness by simply using an ansatz in terms of MPLs $L_w(x)$ of weight $L$, where $w$ is a word of length $L$ with letters 0 or 1. This basis has $2^L$ elements.
We can see that SVMPLs restricted to the real line form a smaller space of functions by noting that the condition $x=\xb$ is a special case of symmetrizing with respect to $z \leftrightarrow \zb$. The symmetrized SVMPLs satisfy
\beq
\cL_w (z) + \cL_w (\zb) = \cL_{\tl w} (z) + \cL_{\tl w} (\zb)\,,
\eeq
where $\tl w$ is the reverse word.
This reduces the number of basis elements of weight $L$ to $2^{L-1} + 2^{\lfloor \frac{L-1}{2}\rfloor}$.
It is also necessary to include in the ansatz products of SVMPLs and the single-valued multiple zeta values (SVMZVs, see \cite{Brown:2013gia})
\beq
\zeta(3)\,, \ 
\zeta(5)\,, \ 
\zeta(3)^2\,, \ \ldots 
\eeq
Including such products, there are
\beq
\sum\limits_{\substack{L\geq 0\\L=W,W-3,W-5,W-6,\ldots}}
2^{L-1} + 2^{\lfloor \frac{L-1}{2}\rfloor}\,,
\eeq
basis elements of weight $W$ that are symmetric under $z \leftrightarrow \zb$.
Next we restrict to the real line, which leads to further relations between the basis elements, starting at weight 4.
We find the relations
\begin{align}
\cL_{0011}(x) ={}& \tfrac{1}{2} \cL_{0110}(x)+\tfrac{1}{2} \cL_{1001}(x) + \zeta (3) \cL_1(x)\,,\nonumber\\
\cL_{01010}(x)={}& 6 \cL_{00011}(x)+2 \cL_{00101}(x)-2 \cL_{00110}(x)-2 \cL_{01001}(x)-3
   \cL_{10001}(x)\,,\nonumber\\
\cL_{10101}(x)={}& 6 \cL_{00111}(x)+2 \cL_{01011}(x)-2 \cL_{01101}(x)-3 \cL_{01110}(x)-2 \cL_{10011}(x)-6 \zeta (3) \cL_{01}(x)\,,\nonumber\\
\cL_{001100}(x)={}&  \cL_{100001}(x)-2 \cL_{000011}(x)+2  \cL_{000110}(x)\nonumber\\
&+2 \zeta (3) \cL_{001}(x)-2 \zeta (3) \cL_{010}(x)+2 \zeta (5) \cL_1(x)\,,\nonumber\\
\cL_{110011}(x)={}&  \cL_{011110}(x)-2   \cL_{001111}(x)+2 \cL_{100111}(x)\nonumber\\
&-2 \zeta (3) \cL_{011}(x)+2 \zeta (3) \cL_{101}(x)+9 \zeta (5) \cL_1(x)\,,\nonumber\\
\cL_{010010}(x)={}&  8 \cL_{000011}(x)+3 \cL_{000101}(x)-2 \cL_{000110}(x)-\cL_{001010}(x)-3 \cL_{010001}(x) \label{svmpl_real_line_relations}\\
&-4 \cL_{100001}(x)-8 \zeta (5) \cL_1(x)\,,\nonumber\\
\cL_{101101}(x)={}&  8 \cL_{001111}(x)+3 \cL_{010111}(x)-3 \cL_{011101}(x)-4 \cL_{011110}(x)-2
   \cL_{100111}(x)\nonumber\\
&-\cL_{101011}(x)-4 \zeta (3) \cL_{011}(x)-2 \zeta (3) \cL_{101}(x)-6 \zeta (5) \cL_1(x)\,,\nonumber\\
\cL_{010101}(x)={}&  9 \cL_{000111}(x)+2 \cL_{001011}(x)-2 \cL_{001101}(x)-3 \cL_{001110}(x)-2 \cL_{010011}(x)\nonumber\\
&-3 \cL_{100011}(x)-3 \zeta (3) \cL_{010}(x)+3 \zeta (3) \cL_{101}(x)+12 \zeta (5) \cL_1(x)\,,\nonumber\\
\cL_{011001}(x)={}&  \cL_{001011}(x)+\cL_{001101}(x)+\cL_{010011}(x)-\cL_{010110}(x)-\cL_{100101}(x)\nonumber\\
&+3 \zeta (3) \cL_{010}(x)-2 \zeta (3) \cL_{011}(x)-3 \zeta (3)
   \cL_{101}(x)-11 \zeta (5) \cL_1(x)\,, \qquad \text{for }x=\xb\,.\nonumber
\end{align}
We show the number of remaining independent basis elements in Table \ref{table:basis_dimensions} and compare it to the number of independent basis elements when using a more general multi-valued basis of MPLs and MZVs. The multi-valued ansatz is enough to fully fix $G^{(1)} (S,T,x)$, as done in \cite{Alday:2024yax}, but not for $G^{(2)} (S,T,x)$.

\begin{table}[h!]
\begin{center}
\begin{tabular}{ l|c|c|c|c|c|c|c } 
 weight & 0 & 1 & 2 & 3 & 4 & 5 & 6\\ 
 \hline
multi-valued  & 1 & 2 & 5 & 11 & 23 & 48 & 98 \\ 
single-valued & 1 & 2 & 3 & 7 & 11 & 22 & 39 \\ 
\end{tabular}
\caption{Number of independent MPLs ($\times$ MZVs) and SVMPLs ($\times$ SVMZVs) on the real line.}
\label{table:basis_dimensions}
\end{center}
\end{table}
In order to implement crossing symmetry we choose a basis that is explicitly (anti-) symmetric under $x \leftrightarrow 1-x$, writing
\beq
\cL^\pm_w (x) = \cL_w (x) \pm \cL_w (1-x)\,.
\eeq
To relate this to the previous discussion it is useful to know that
\beq
\cL_w (1-z) = \cL_{\hat w} (z) + \text{ SVMZVs times lower weight SVMPLs}\,,
\eeq
where $\hat{w}$ is the word where each letter is flipped $(0 \leftrightarrow 1)$.
For example
\beq
\cL_{011} (1-z) = \cL_{100} (z) + 2 \zeta(3)\,.
\eeq
Our choice of basis for SVMPLs on the real line up to weight 6 is the following
\begin{align}
T^+_{0}(x) ={}& \left(1  \right)\,, \qquad
T^+_{1}(x) = \left(\cL^+_{0}(x)  \right)\,, \qquad
T^-_{1}(x) = \left(\cL^-_{0}(x) \right)\,,\nonumber\\
T^+_{2}(x) ={}& \left(\cL^+_{00}(x), \cL^+_{01}(x)  \right)\,, \qquad
T^-_{2}(x) = \left(\cL^-_{00}(x)\right)\,,\nonumber\\
T^+_{3}(x) ={}& \left(\cL^+_{000}(x), \cL^+_{001}(x), \cL^+_{010}(x), \zeta(3) \right)\,,\nonumber\\
T^-_{3}(x) ={}& \left(\cL^-_{000}(x), \cL^-_{001}(x), \cL^-_{010}(x)  \right)\,,\nonumber\\
T^+_{4}(x) ={}& \left(\cL^+_{0000}(x),\cL^+_{0001}(x),\cL^+_{0010}(x),\cL^+_{0101}(x),\cL^+_{0110}(x),\zeta (3) \cL^+_0(x) \right)\,,\nonumber\\
T^-_{4}(x) ={}& \left(\cL^-_{0000}(x),\cL^-_{0001}(x),\cL^-_{0010}(x),\cL^-_{0110}(x),\zeta (3) \cL^-_0(x) \right)\,,\nonumber\\
T^+_{5}(x) ={}& \big(\cL^+_{00000}(x),\cL^+_{00001}(x),\cL^+_{00010}(x),\cL^+_{00011}(x),\cL^+_{00100}(x),\cL^+_{00101}(x),\nonumber\\
&\ \,\cL^+_{00110}(x),\cL^+_{01001}(x),\cL^+_{01110}(x),\zeta (3) \cL^+_{00}(x),\zeta (3) \cL^+_{01}(x),\zeta (5) \big)\,,\label{MPL_list}\\
T^-_{5}(x) ={}& \big(\cL^-_{00000}(x),\cL^-_{00001}(x),\cL^-_{00010}(x),\cL^-_{00011}(x),\cL^-_{00100}(x),\cL^-_{00101}(x),\nonumber\\
&\ \,\cL^-_{00110}(x),\cL^-_{01001}(x),\cL^-_{01110}(x),\zeta (3) \cL^-_{00}(x) \big)\,,\nonumber\\
T^+_{6}(x) ={}& \big(\cL^+_{000000}(x),\cL^+_{000001}(x),\cL^+_{000010}(x),\cL^+_{000011}(x),\cL^+_{000100}(x),\cL^+_{000101}(x),\cL^+_{000110}(x),\nonumber\\
&\ \,\cL^+_{000111}(x),\cL^+_{001001}(x),\cL^+_{001010}(x),\cL^+_{001011}(x),\cL^+_{001101}(x),\cL^+_{001110}(x),\cL^+_{010001}(x),\nonumber\\
&\ \,\cL^+_{010110}(x),\cL^+_{011110}(x),\zeta (3) \cL^+_{000}(x),\zeta (3)
   \cL^+_{001}(x),\zeta (3) \cL^+_{010}(x),\zeta (5) \cL^+_0(x),\zeta (3)^2 \big)\,,\nonumber\\
T^-_{6}(x) ={}& \big(\cL^-_{000000}(x),\cL^-_{000001}(x),\cL^-_{000010}(x),\cL^-_{000011}(x),\cL^-_{000100}(x),\cL^-_{000101}(x),\cL^-_{000110}(x),\nonumber\\
&\ \,\cL^-_{001001}(x),\cL^-_{001010}(x),\cL^-_{001101}(x),\cL^-_{001110}(x),\cL^-_{010001}(x),\cL^-_{010110}(x),\cL^-_{011110}(x),\nonumber\\
&\ \,\zeta (3) \cL^-_{000}(x),\zeta (3) \cL^-_{001}(x),\zeta (3)
   \cL^-_{010}(x),\zeta (5) \cL^-_0(x) \big)\,.\nonumber
\end{align}

\subsection{Ansatz}

We make the following ansatz for the $k$th curvature correction to the world-sheet correlator
\beq
G^{(k)} (S,T,x) =  \frac{1}{(S+T)^k} \sum\limits_{n=0}^{3k} \sum\limits_{j,\pm}  P^{(k)\pm}_{n,j}(S,T) T^\pm_{n,j}(x)\,,
\label{Gk_ansatz}
\eeq
where $P^{(k)\pm}_{n,j}(S,T)$ are symmetric / antisymmetric homogeneous polynomials of degree $n$ with rational coefficients.
The sum runs over the independent SVMPLs on the real line \eqref{MPL_list} up to weight $3k$.

We would like to compare the ansatz to \eqref{A_dr_poles}. To this end one first Taylor expands the integrand of \eqref{worldsheet_integral} around $x=0$ and then integrates before finally expanding around $S=\delta$, for different values of $\delta$.
The matching implies that the OPE data appearing in \eqref{A_dr_poles} can only depend on MZVs that explicitly appear in the world-sheet integrand, i.e.\ SVMZVs. For $k=1$ we saw already in \cite{Alday:2024yax} that the OPE data has the form
\bea
\sqrt{\delta} \langle f_0 \tau_2 \rangle_{\delta, \ell} &= r^{1}_{\delta,\ell} \,,\\
\langle f_2 \rangle_{\delta, \ell} &= r^{2}_{\delta,\ell} +  r^{3}_{\delta,\ell} \zeta(3)\,,
\eea{OPE_assumption_1}
where $r^{i}_{\delta,\ell}$ are rational numbers.
For $k=2$ we make the ansatz
\begin{align}
\langle f_0 \tau_2^2 \rangle_{\delta, \ell} ={}& r^{4}_{\delta,\ell} \,,\nonumber\\
\sqrt{\delta} \langle f_0 \tau_4 + f_2 \tau_2 \rangle_{\delta, \ell} ={}&
r^{5}_{\delta,\ell} 
+ r^{6}_{\delta,\ell} \zeta(3)\,,\label{OPE_assumption_2}\\
\langle f_4 \rangle_{\delta, \ell} ={}&
r^{7}_{\delta,\ell} 
+ r^{8}_{\delta,\ell} \zeta(3)
+ r^{9}_{\delta,\ell} \zeta(5)
+ r^{10}_{\delta,\ell} \zeta(3)^2\,.
\nonumber
\end{align}
In general we expect $\langle f_{2k} \tau_2^{j_2} \tau_4^{j_4} \tau_6^{j_6} \ldots \rangle_{\delta, \ell}$ to have terms of maximal weight $3(k + j_4 + 2 j_6 + \ldots)$.

\subsection{Solution}
The ansatz for $k=2$ has $254$ rational coefficients, of which all but one are fixed by matching the ansatz with the dispersion relation.
The polynomials appearing in the solution are given by\footnote{We also include the final expression for $G^{(2)} (S,T,x)$ in an ancillary Mathematica file.}
\begin{align}
{}&P^{(2)+}_{0}(S,T) = \big( \tfrac{19}{2} \big)\,, \qquad
P^{(2)+}_{1}(S,T) = (S+T)\big( \tfrac{19}{8} \big)\,, \qquad
P^{(2)-}_{1}(S,T) = (T-S)\big( \tfrac{117}{8} \big)\,, \nonumber\\
&P^{(2)+}_{2}(S,T) = (S+T)^2 \big(\tfrac{67}{8},-\tfrac{29}{4}  \big) + S T \big(-\tfrac{109}{4},\tfrac{109}{4}  \big)\,, \qquad
P^{(2)-}_{2}(S,T) = (T^2 - S^2) \big( \tfrac{61}{8}  \big)\,, \nonumber\\
&P^{(2)+}_{3}(S,T) = (S+T)^3 \big( -\tfrac{21}{4},\tfrac{3}{2},\tfrac{1}{4},\tfrac{7}{2}\big)
+ S T(S+T) \big( -\tfrac{97}{4},\tfrac{71}{4},\tfrac{13}{2},0\big)\,, \nonumber\\
&P^{(2)-}_{3}(S,T) = (S-T)(S+T)^2 \big(-\tfrac{21}{4},\tfrac{7}{4},\tfrac{1}{2} \big)
+ (S-T)S T \big(\tfrac{21}{4},-\tfrac{21}{2},-\tfrac{21}{4} \big)\,, \nonumber\\
&P^{(2)+}_{4}(S,T) = (S+T)^4 \big( \tfrac{23}{2},-\tfrac{95}{8},-12,\tfrac{55}{8},\tfrac{25}{2},2\big)
+ S T(S+T)^2 \big(-\tfrac{65}{4},\tfrac{39}{8},\tfrac{3}{8},\tfrac{27}{8},\tfrac{61}{8},-\tfrac{7}{4} \big)\nonumber\\
&+ S^2 T^2 \big( \tfrac{3}{2},-3,-3,\tfrac{3}{2},3,0\big)\,, \nonumber\\
&P^{(2)-}_{4}(S,T) = (S-T)(S+T)^3 \big(\tfrac{23}{2},-\tfrac{95}{8},-12,-\tfrac{5}{8},2 \big)
+S T (S^2-T^2) \big(\tfrac{27}{4},-\tfrac{63}{8},-\tfrac{45}{8},-\tfrac{9}{8},\tfrac{9}{4} \big)\,, \nonumber\\
&P^{(2)+}_{5}(S,T) = (S+T)^5 \big(-\tfrac{27}{4},\tfrac{15}{2},\tfrac{75}{8},-\tfrac{39}{2},\tfrac{39}{8},-\tfrac{77}{8},\tfrac{1}{2},\tfrac{11}{8},\tfrac{27}{4},-\tfrac{7}{4},-\tfrac{43}{4},-\tfrac{165}{2} \big)\nonumber\\
&+ S T(S+T)^3 \big(19,-\tfrac{75}{4},-\tfrac{81}{4},\tfrac{61}{4},-\tfrac{41}{4},\tfrac{41}{4},-\tfrac{3}{4},\tfrac{17}{4},\tfrac{5}{4},5,\tfrac{9}{2},9 \big)\label{G2_solution}\\
&+ S^2 T^2(S+T) \big(\tfrac{5}{2},-\tfrac{7}{2},-\tfrac{11}{4},\tfrac{11}{4},-\tfrac{5}{4},\tfrac{3}{2},-\tfrac{1}{4},\tfrac{3}{4}
   ,\tfrac{1}{4},\tfrac{1}{2},0,-3 \big)\,, \nonumber\\
&P^{(2)-}_{5}(S,T) = (S-T)(S+T)^4 \big(-\tfrac{27}{4},\tfrac{15}{2},\tfrac{75}{8},-\tfrac{39}{2},\tfrac{39}{8},-\tfrac{77}{8},\tfrac{1}{2},\tfrac{11}{8},-\tfrac{27}{4},-\tfrac{7}{4} \big)\nonumber\\
&+(S-T)S T (S+T)^2 \big( \tfrac{11}{2},-\tfrac{15}{4},-\tfrac{3}{2},-\tfrac{35}{4},-\tfrac{1}{2},-4,-\tfrac{3}{4},\tfrac{1}{2},-\tfrac{11}{4},\tfrac{3}{2}\big)\,, \nonumber\\
&P^{(2)+}_{6}(S,T) = (S+T)^6 \big(\tfrac{5}{4},-\tfrac{5}{4},-\tfrac{7}{4},\tfrac{11}{4},-2,\tfrac{15}{8},1,0,\tfrac{5}{8},\tfrac{1}{2},0,0,0, -\tfrac{3}{4},0,-1,\tfrac{1}{4},\tfrac{7}{4},-\tfrac{1}{4},0,\tfrac{9}{2} \big)\nonumber\\
&- S T(S+T)^4 \big(5,-5,-6,10,-\tfrac{13}{2},\tfrac{25}{4},2,-\tfrac{27}{4},\tfrac{5}{2},\tfrac{5}{4},-\tfrac{11}{4},\tfrac{1}{4},\tfrac{15}{4},-\tfrac{3}{2},\tfrac{1}{2},-3,1,2,2,0,8 \big)\nonumber\\
&+ S^2 T^2(S+T)^2 \big(\tfrac{5}{2},-\tfrac{5}{2},-\tfrac{3}{2},-\tfrac{1}{2},-1,-\tfrac{3}{4},-\tfrac{3}{2},\tfrac{19}{4},-\tfrac{1}{4},-\tfrac{1}{2},2,\tfrac{1}{4},-\tfrac{7}{4},\tfrac{1}{2},-\tfrac{1}{4},\tfrac{1}{2},\tfrac{1}{2},1,-\tfrac{1}
   {2},0,2 \big)\,, \nonumber\\
&P^{(2)-}_{6}(S,T) = (S-T)(S+T)^5 \big(\tfrac{5}{4},-\tfrac{5}{4},-\tfrac{7}{4},\tfrac{11}{4},-2,\tfrac{15}{8},1,\tfrac{5}{8},\tfrac{1}{2},0,0,-\tfrac{3}{4},0,1,\tfrac{1}{4},\tfrac{7}{4},-\tfrac{1}{4},0 \big)\nonumber\\
&+(S-T)S T (S+T)^3 \big(-\tfrac{5}{2},\tfrac{5}{2},\tfrac{5}{2},-\tfrac{9}{2},\tfrac{5}{2},-\tfrac{5}{2},0,-\tfrac{5}{4},-\tfrac{1}{4},-\tfrac{1}{4},-\tfrac{3}{4},0,-\tfrac{1}{2},-1,-\tfrac{1}{2},-\tfrac{3}{2},\tfrac{1}{2},0 \big)\,,\nonumber
\end{align}
where we fixed the final coefficient to the correct value for the $SO(8)$ theory, as will be described in section \ref{sec:LEE} below.
Keeping this coefficient, let us call it $c^{(2)}$, unfixed corresponds to adding the following to $G^{(2)} (S,T,x)$
\begin{align}
{}&c^{(2)} \bigg[\frac{6 S^2+7 S T+6 T^2}{16}  \cL^+_{0101}(x)
-\frac{S^2+4 S T+T^2}{16}  \cL^+_{0010}(x)
-\frac{5 S^2+3 S T+5 T^2}{16}  \cL^+_{0110}(x)\nonumber\\
&+\frac{S^2-4S T+T^2}{8 (S+T)} \left(\cL^+_{010}(x)-\cL^+_{001}(x) \right)
+\frac{5 S^2-2 S T+5 T^2}{16 (S+T)^2} \left( \cL^+_{01}(x) -\cL^+_{00}(x)\right)\nonumber\\
&+ \left(\frac{17}{16}  \left(S^2+T^2\right)\zeta (3)+\frac{1}{4 (S+T)}\right)\cL^+_0(x)
-2 (S+T)\zeta (3) +\frac{1}{(S+T)^2}\nonumber\\
&+(S-T) \bigg(\frac{1}{16} (S+T) \left(\cL^-_{0110}(x)-\cL^-_{0010}(x)\right)
-\frac{3}{8} \cL^-_{001}(x)+\frac{1}{8} \cL^-_{010}(x)-\frac{5 \cL^-_{00}(x)}{16 (S+T)}\nonumber\\
&+ \left(\frac{17}{16}  (S+T)\zeta (3)+\frac{1}{2 (S+T)^2}\right)\cL^-_0(x)\bigg)\bigg]\,.
\label{G2extra}
\end{align}
In the theories with $G_F = U(4)$ or $SO(4) \times SO(4)$ the world-sheet integrand might differ by this term. This question could potentially be resolved by doing the localisation computations for these theories.

\section{Checks}
\label{sec:checks}

In obtaining our solution we made several assumptions, so it is important to perform consistency checks.
In this section we show that the high energy limit of $A^{(2)}(S,T)$ is consistent with exponentiation as expected from a classical scattering computation \cite{Alday:2023pzu},
the low energy expansion is free of poles at $S,T=0$ and the OPE data matches the results for the energy of classical solutions for massive string operators.
We also fully fix the $D^6 F^4$ correction to the Mellin amplitude and provide a wealth of OPE data that can be compared to future computations with other methods.

\subsection{High energy limit}

In the high-energy limit of large $S,T,R$ with $S/T$ and $S/R$ fixed\footnote{We set $\alpha'=1$ in this section.} we expect
the amplitude $A(S,T)$ to be determined by a classical computation, as shown for the closed string amplitude on $AdS_5 \times S^5$ in \cite{Alday:2023pzu}.
We expect the form
\beq
A^{\text{HE}}(S,T) \equiv
\lim\limits_{\substack{S,T,R \to \infty\\ S/T, S/R \text{ fixed}}}
A(S,T) \sim
e^{-\mathcal{E}_\text{open}(S,T)}\,,
\label{HE_limit}
\eeq
and the exponent is determined by the saddle point at $x=\frac{S}{S+T}$ of the integral \eqref{worldsheet_integral} for $k=0,1$
\begin{align}
\mathcal{E}_\text{open}(S,T) ={}& \mathcal{E}^{(0)}(S,T) + \frac{1}{R^2} \mathcal{E}^{(1)}(S,T) + O \left(\frac{1}{S} \right)\,,\nonumber\\
\mathcal{E}^{(0)}(S,T) ={}& S \log \left(\tfrac{S}{S+T}\right) + T \log \left(\tfrac{T}{S+T}\right)\,,\nonumber\\
\mathcal{E}^{(1)}(S,T) ={}& - G^{(1)} \left(S,T,\tfrac{S}{S+T}\right)+ O(S)
 \label{Eopen}\\
={}&\frac{S^2}{2} \left(\cL_{001}\left( \tfrac{S}{S+T} \right)- \cL_{000}\left( \tfrac{S}{S+T} \right) -2\zeta (3)\right)
   +\frac{S (S+T)}{2} \cL_{010}\left( \tfrac{S}{S+T} \right)\nonumber\\
&+\frac{T^2}{2} \left( \cL_{110}\left( \tfrac{S}{S+T} \right)- \cL_{111}\left( \tfrac{S}{S+T} \right)\right)
   +\frac{T (S+T)}{2} \cL_{101}\left( \tfrac{S}{S+T} \right)\,.
\nonumber
\end{align}
Crucially, the exponent depends only on the first curvature correction (see \cite{Alday:2023pzu}). This means that the high energy limit of all further curvature corrections is determined by \eqref{HE_limit}. In particular
\beq
A^{\text{HE}}(S,T)\sim
e^{-\mathcal{E}^{(0)}(S,T)} \left(1 - \frac{1}{R^2} \mathcal{E}^{(1)}(S,T) + \frac{1}{2R^4} \left(\mathcal{E}^{(1)}(S,T)\right)^2+  \ldots\right)\,,
\eeq
which implies
\beq
G^{(2)} (S,T,\tfrac{S}{S+T}) = \frac12 \left(\mathcal{E}^{(1)}(S,T)\right)^2 + O(S^3)\,,
\label{exponentiation}
\eeq
which we confirm to be true for our solution.

Furthermore, combining \eqref{exponentiation} with the relation between the high energy exponents for open and closed strings \cite{Alday:2024yax}
\beq
\mathcal{E}_\text{open}(S,T) = \frac12 \mathcal{E}_\text{closed}(4S,4T)\,,
\eeq
implies an identity between the high energy limits of $A^{(2)}(S,T)$ for open and closed strings.

\subsection{Low energy expansion}
\label{sec:LEE}

Let us now discuss the low energy expansion of $A(S,T)$, i.e.\ the Taylor expansion around $S = T = 0$.
We define the Wilson coefficients $\alpha^{(k)}_{a,b}$ by
\beq
A^{(k)}(S,T) =  -\frac{\delta_{k,0}}{ST}+ \sum\limits_{a,b=0}^\infty 
\hat\sigma_1^a
\hat\sigma_2^b
 \alpha^{(k)}_{a,b} \,,
\qquad
\hat\sigma_1= -U\,, \quad
\hat\sigma_2 = -ST\,.
\label{LEE}
\eeq
To compute this expansion from the world-sheet integral \eqref{worldsheet_integral}, we first express the integrand in terms of usual MPLs and consider integrals of the form
\beq
I_w (S,T) = \int\limits_0^1 dx \, x^{-S-1} (1-x)^{-T-1} L_w (x)\,.
\eeq
The expansion of these integrals was done in \cite{Alday:2024yax}, with the result
\beq
I_w (S,T) = \text{poles } + 
\sum\limits_{p,q=0}^{\infty} (-S)^p (-T)^q 
\sum\limits_{W \in 0^p \shuffle 1^q \shuffle w} 
\left( L_{0W}(1) - L_{1W}(1) \right)\,,
\eeq
where $\shuffle$ is the shuffle product.
Expanding the result \eqref{G2_solution} including the unfixed term \eqref{G2extra} gives
\beq
A^{(2)}(S,T) = \left(48-\frac{3 c^{(2)}}{2}\right) \zeta (2)^2
+\left(\frac{1851}{4}-\frac{69 c^{(2)}}{8}\right) \zeta (5) \hat\sigma_1 + \ldots
\eeq
For the $G_F = SO(8)$ theory the leading coefficient in this expansion was computed using supersymmetric localisation in \cite{Behan:2023fqq,Alday:2024yax} to be
\beq
\alpha^{(2)}_{0,0} = 48 \zeta(2)^2\,,
\label{localisation}
\eeq
so indeed we have to set $c^{(2)} = 0$ for this theory, which we will do for most of the remainder of the paper. The low energy expansion for $A^{(2)}(S,T)$ is then
\bea
A^{(2)}(S,T) ={}& 48 \zeta (2)^2
+\hat\sigma _1\tfrac{1851}{4} \zeta (5)
+\hat\sigma _1^2   \left(\tfrac{53128}{105} \zeta (2)^3-125 \zeta (3)^2\right)\\
&+\hat\sigma _2 \left(\tfrac{88288 }{105}\zeta (2)^3+250 \zeta (3)^2\right)
+\hat\sigma _1^3 \left(\tfrac{2016}{5} \zeta (3) \zeta (2)^2+\tfrac{174081 }{32}\zeta (7)\right)\\
&+ \hat\sigma _1\hat\sigma _2 \left(\tfrac{12036}{5}   \zeta (3) \zeta (2)^2+\tfrac{16503}{4} \zeta (5) \zeta (2)+\tfrac{174081 }{32}\zeta (7)\right)\\
&+\hat\sigma _1^4 \left(\tfrac{12016819}{5250} \zeta (2)^4+1384 \zeta (3)^2 \zeta
   (2)-\tfrac{5447}{4} \zeta (3) \zeta (5) + \tfrac{33}{20} \zeta (5,3)\right)\\
&+\hat\sigma _1^2\hat\sigma _2  \left(\tfrac{4057994}{525} \zeta (2)^4+2768 \zeta (3)^2 \zeta (2)+\tfrac{32657}{4} \zeta (3) \zeta (5)-\tfrac{57}{4} \zeta(5,3)\right)\\
&+\hat\sigma _2^2 \left(\tfrac{10723856}{2625} \zeta (2)^4+1384 \zeta (3)^2 \zeta (2)+\tfrac{5447}{2} \zeta (3) \zeta (5)-\tfrac{33}{10} \zeta (5,3)\right)
+\ldots\,.
\eea{A2_LEE2}
This new result allows us to completely fix the Mellin amplitude up to order $\lambda^{-5/2}$, which corresponds to the $D^6 F^4$ term in the low energy effective action. The low energy expansion of the Mellin amplitude is related to \eqref{LEE} by \eqref{borel}
\beq
M(s,t) = -\frac{2}{s \, t} +  \sum\limits_{k,a,b=0}^\infty \frac{\Gamma(4+a+2b)2^{3+a+2b}}{\lambda^{1+\frac{a}{2}+b+\frac{k}{2}}}
\sigma_1^a
\sigma_2^b
 \alpha^{(k)}_{a,b} \,,
\label{wilson_expansion}
\eeq
with
\beq
\sigma_1 = -u\,, \qquad
\sigma_2 = -s t\,.
\eeq
The first few terms of the Mellin amplitude read \cite{Behan:2023fqq,Alday:2024yax}
\begin{align}
M(s,t) &= -\frac{2}{s \, t}+\frac{48 \zeta (2)}{\lambda }+ \frac{384\zeta(3) \sigma _1}{\lambda^\frac32 }  +\frac{384 \zeta (2)^2 \left(4 \sigma _1^2+7 \sigma_2-3 \sigma _1+6\right) }{\lambda^2}\label{mellin_lee}\\
& + \frac{1}{\lambda^\frac52}
\left(960 \left( \alpha^{(0)}_{3,0} \sigma _1^3 + \alpha^{(0)}_{1,1} \sigma _1 \sigma _2 
\right)
+80 \left( \alpha^{(1)}_{2,0} \sigma _1^2 +\alpha^{(1)}_{0,1} \sigma _2 \right)
+8  \alpha^{(2)}_{1,0} \sigma _1 +\alpha^{(3)}_{0,0} \right) +O\left( \lambda^{-3} \right)\,.\nonumber
\end{align}
We can extract most of the coefficients at $O(\lambda^{-5/2})$ from the known $A^{(k=0,1,2)}(S,T)$
\bea
\alpha^{(0)}_{3,0} &= \zeta(5) \,, &\qquad
\alpha^{(0)}_{1,1} &= \zeta(5) + \zeta(2) \zeta(3) \,,\\
\alpha^{(1)}_{2,0} &= \frac{19}{2}\zeta(5) - 20 \zeta(2) \zeta(3) \,, &\qquad
\alpha^{(1)}_{0,1} &= -19\zeta(5) -20 \zeta(2) \zeta(3)\,, \qquad \\
\alpha^{(2)}_{1,0} &= \frac{1851}{4} \zeta(5) \,. &&
\eea{wilson52}
The final coefficient can then be fixed using the localisation constraint of \cite{Behan:2023fqq} (see also appendix C of \cite{Alday:2024yax})
\beq
\alpha^{(3)}_{0,0} = 64\alpha^{(1)}_{2,0} + 32 \alpha^{(1)}_{0,1}
= -1920 \zeta(2) \zeta(3)  \,.
\eeq

\subsection{OPE data}

Let us now extract the OPE data from our solution. From $A^{(2)}(S,T)$ we obtain for the first Regge trajectory\footnote{Here we set $c^{(2)} = 0$. We include the dependence on $c^{(2)}$ and analogous formulas for further Regge trajectories in an ancillary Mathematica file.}
\begin{align}
( f_0 \tau_2^2 )_{\delta, \delta-1} ={}& r_0(\delta) \delta\left(\frac{3\delta}{4}  +\frac{1}{2\delta }-\frac{3}{4}\right)^2\,,\nonumber\\
( f_0 \tau_4 + f_2 \tau_2 )_{\delta, \delta-1} ={}&
r_0(\delta) \sqrt{\delta} \left(
-\frac{7 \delta ^3}{16}+\frac{191 \delta ^2}{32}+\frac{79}{16 \delta ^2}-\frac{1217 \delta
   }{48}-\frac{1853}{96 \delta }+\frac{249}{8}\right)\nonumber\\
&+\delta^2 \zeta(3) ( f_0 \tau_2 )_{\delta, \delta-1} - \frac{7\delta^\frac32}{2 }  \zeta(3) ( f_0 )_{\delta, \delta-1} 
\,, \label{A2_OPE_data_0}\\
( f_4 )_{\delta, \delta-1} ={}&
r_0(\delta) \bigg(
\frac{49 \delta ^4}{288}-\frac{253 \delta ^3}{80}+\frac{5563 \delta ^2}{288}+\frac{1611}{128 \delta
   ^2}-\frac{8909 \delta }{96}-\frac{6889}{60 \delta }+\frac{54823}{288}\nonumber\\
&-\left(\frac{7 \delta ^4}{12}-\frac{27 \delta^3}{4}+\frac{104 \delta ^2}{3}-\frac{45 \delta }{8}\right) \zeta (3)
\bigg) + \left(\frac{\delta^4}{2} \zeta(3)^2 - \frac{3 \delta^3}{2} \zeta(5) \right)
( f_0 )_{\delta, \delta-1} \nonumber
\end{align}
where
\beq
r_n(\delta) = \frac{4^{1-\delta} \delta^{\delta-2n-1} (\delta-2n)^2}{\Gamma(\delta-n+1)}\,.
\label{r}
\eeq
For the second Regge trajectory we find
\begin{align}
\langle f_0 \tau_2^2 \rangle_{\delta, \delta-3} ={}& r_1(\delta) \left(\frac{3 \delta ^5}{16}+\frac{\delta ^4}{6}-\frac{479 \delta ^3}{432}+\frac{2119 \delta ^2}{216}-\frac{341
   \delta }{36}+\frac{2}{3}\right)\,,\nonumber\\
\langle f_0 \tau_4 + f_2 \tau_2 \rangle_{\delta, \delta-3} ={}&
r_1(\delta) \sqrt{\delta} \left(-\frac{7 \delta ^5}{48}+\frac{679 \delta ^4}{864}-\frac{2293 \delta ^3}{216}-\frac{226 \delta
   ^2}{9}+\frac{8761 \delta }{96}+\frac{725}{12 \delta }-\frac{7031}{72}\right)\nonumber\\
&+\delta^2 \zeta(3) \langle f_0 \tau_2 \rangle_{\delta, \delta-3} - \frac{7\delta^\frac32}{2 }  \zeta(3) \langle f_0 \rangle_{\delta, \delta-3} 
\,, \nonumber\\
\langle f_4 \rangle_{\delta, \delta-3} ={}&
r_1(\delta) \bigg(\frac{49 \delta ^6}{864}-\frac{913 \delta ^5}{3240}+\frac{25927 \delta ^4}{4320}+\frac{116401 \delta^3}{2592}+\frac{182401 \delta ^2}{2592}+\frac{75427 \delta }{2160}\nonumber\\
&+\frac{317771}{192 \delta}-\frac{2843297}{1920}-\left(\frac{7 \delta ^6}{36}-\frac{79 \delta ^5}{108}+\frac{929 \delta ^4}{54}+\frac{847 \delta^3}{24}-\frac{877 \delta ^2}{12}\right) \zeta (3)\bigg)\nonumber\\
& + \left(\frac{\delta^4}{2} \zeta(3)^2 - \frac{3 \delta^3}{2} \zeta(5) \right)
\langle f_0 \rangle_{\delta, \delta-3} \,.
\label{A2_OPE_data_1}
\end{align}
Note that the angle brackets indicate a sum over operators with the same $(\delta,\ell)$.
The operators on the first Regge trajectory are non-degenerate so that we can combine the data with the one at lower orders \cite{Alday:2024yax} and solve for the dimensions $\Delta(\delta,\ell) \equiv \tau(\delta,\ell)+\ell$
\begin{align}
\label{Delta}
{}&\Delta(\delta,\delta-1) = \sqrt{\delta} \lambda^{\frac14}
\Bigg[1+ \frac{1}{\sqrt{\lambda}} \left( \frac{3\delta}{4} + \frac{1}{2\delta} - \frac34 \right)\\
&+\frac{1}{\lambda} \left(-\frac{21 \delta ^2}{32}-\frac{1}{8 \delta ^2}-\frac{(3+14 \zeta(3)) \delta }{4}+\frac{3}{8 \delta }+\frac{41}{32}
+\left(  \frac{(19+34 \zeta (3))\delta}{16}-\frac{11}{16} \right) c^{(2)}
\right) + O\left(\frac{1}{\lambda^{\frac32}}\right)\Bigg]
\nonumber
\end{align}
This can be compared to the classical glued folded open string solution in AdS as discussed in \cite{Alday:2024yax}, which has the energy
\bea
        E ={}&  \sqrt{\delta} \lambda^{\frac14}
 \Bigg[1 + \frac{1}{\sqrt{\lambda} } \left(\frac{3\delta}{4} +\frac{1}{2\delta} - \frac{1}{2} + a^{(1)} \right) + \frac{1}{\lambda } \left(-\frac{21 \delta^2}{32}-\frac{1}{8 \delta^2}+ b_0^{(1)}  \delta+  \frac{b_1^{(1)}}{\delta}+ b^{(2)} \right)  + \ldots \Bigg]\,.
\eea{Equant}
While the coefficients $a^{(1)}, b_0^{(1)}, b_1^{(1)}$ and $b^{(2)}$ are determined by the 1-loop and 2-loop fluctuations around the classical solution, and are currently unavailable from this perspective, we find an exact match for all the classical terms.

\section{Conclusions}
\label{sec:conclusions}

In this paper we argued that the world-sheet integrand of the AdS Veneziano amplitude in a small curvature expansion should be expressible in terms of single-valued multiple polylogarithms evaluated on the real axis,
because curvature corrections correspond to soft gravitons,
which are governed by single-valued insertions.

Combining this with a CFT dispersion relation we fixed the second curvature correction
up to one rational coefficient $c^{(2)}$.
For the theory with flavour group $G_F=SO(8)$ this coefficient can be fixed using the localisation constraint computed in \cite{Behan:2023fqq}.
For the other possibilities $G_F = U(4)$ or $SO(4)\times SO(4)$
the localisation constraint has not been computed yet, so $c^{(2)}$ is not fixed.
Another way to fix it would be to compute quantum corrections to the energy of the glued folded open string solution in AdS and compare it to our result for the conformal dimensions of massive string operators \eqref{Delta}.
At the classical level, the two quantities already agree.
A further consistency check is exponentiation in the high energy limit.
Our result also fully fixes the unprotected $D^6 F^4$ term in the low energy effective action.

The results of this paper lead to a new way of thinking about KLT \cite{Kawai:1985xq}/double-copy \cite{Bern:2008qj,Bern:2010ue} type relations on curved backgrounds. The scattering of open strings in a curved background should be thought of as the scattering of open strings with extra gravitons.
If there are to be any double-copy relations, it seems that only the open strings, but not the soft gravitons from curvature corrections, should be doubled.
The scattering of these extra gravitons is already governed by single-valuedness, restricted to the real line in the context of open string scattering. 

Finally, we came across
single-valued multiple polylogarithms evaluated on the real line.
These also appear in different physical applications, for instance in the context of Wilson line defect CFTs, see \cite{Ferrero:2023gnu}, but to our knowledge these functions have not been studied systematically. It would be interesting to study this from a mathematical perspective.

\section*{Acknowledgements}

We thank Erik Panzer for useful discussions.
The work of LFA is supported by the European Research Council (ERC) under the European Union's Horizon 2020 research and innovation programme (grant agreement No 787185). LFA is also supported in part by the STFC grant ST/T000864/1.
TH is supported by the STFC grant ST/X000591/1.

\appendix

\section{Computation of ${\cal J}(x,\bar x)$}
\label{app:J}
We would like to compute 
\begin{equation}
{\cal J}(x,\bar x) = \int_\mathbb{C} d^2z \frac{|z|^{2a}|1-z|^{2b}|z-x|^{2c}}{|z|^2|1-z|^2}\,,
\end{equation}
in a small $a,b,c$ expansion, and show that the resulting functions are single-valued functions of $x$. Following \cite{Vanhove:2020qtt} we write this in a factorised form
\begin{equation}
{\cal J}(x,\bar x) =-\frac{1}{\pi} \left(\kappa_1 J_1(x) J_1(\bar x)+ \kappa_2 J_2(x) J_2(\bar x)\right)\,,
\end{equation}
with $\kappa_1 = \sin(\pi a) \csc(\pi (b + c)) \sin(\pi(a+b+c)),\kappa_2=\sin(\pi b) \sin(\pi c) \csc(\pi(b+c))$ and 
\bea
J_1(x) &= \int_{-\infty}^0 (-z)^{a+1}(1-z)^{b-1}(x-z)^c dz\,,\\
J_2(x) &= \int_{x}^1 (z)^{a+1}(1-z)^{b-1}(z-x)^c dz\,.
\eea{J12}
These integrals can be evaluated in terms of linear combinations of the following two hypergeometric functions
\begin{equation}
h_1(x) = \, _2F_1(a,1-b;a+c+1;x)\,,~~~h_2(x)=\, _2F_1(-a-b-c+1,-c;-a-c+1;x)\,,
\end{equation}
which leads to the following expression for ${\cal J}(x,\bar x)$
\begin{equation}
{\cal J}(x,\bar x) =\hat \kappa_1 |x|^{2a+2c} h_1(x) h_1(\bar x) +\hat \kappa_2 h_2(x) h_2(\bar x) \,,
\end{equation}
with
\bea
\hat \kappa_1&=-\frac{\pi  \Gamma (-a-c)^2 \csc (\pi  (b+c)) (\csc (\pi  a) \sin (\pi  (a+b+c))+\sin (\pi  b) \csc (\pi  c))}{\Gamma (1-a)^2 \Gamma (-c)^2}\,,\\
\hat \kappa_2&= -\frac{\pi  \Gamma (a+c)^2 \csc (\pi  (b+c)) (\sin (\pi  a) \csc (\pi  (a+b+c))+\csc (\pi  b) \sin (\pi  c))}{\Gamma (1-b)^2 \Gamma (a+b+c)^2}\,.
\eea{kappa12}
We now note the following. First, $\hat \kappa_1,\hat \kappa_2$ admit an expansion around small $(a,b,c)$ with coefficients proportional to SVMZVs. Schematically
\beq
\hat \kappa_1,\hat \kappa_2 \sim \frac{1}{\epsilon} + \zeta(3) \epsilon^2 + \zeta(5) \epsilon^4+ \zeta(3)^2 \epsilon^5+ \cdots\,,
\eeq
with $(a,b,c) \sim \epsilon$. Second, $h_1(x), h_2(x)$ are both of the form 
\beq
h(x) =  \, _2F_1(1+\epsilon_1,\epsilon_2;1+\epsilon_3;x)\,.
\eeq
These hypergeometric functions admit an expansion around small $\epsilon_i$ of the form
\beq
h(x) = 1+ \epsilon h^{(1)}(x) + \epsilon^2  h^{(2)}(x)+\cdots\,,
\eeq
where $h^{(k)}(x)$ is a linear combination of multiple polylogarithms $L_w(x)$ of weight $k$, labelled by words in the alphabet $\{0,1 \}$, analytic (and also vanishing) at $x=0$, so that their last letter is always $1$. The functions $h^{(k)}(x)$ can be recursively computed as follows. First we write $h^{(k)}(x) $ as a linear combination of MPLs of weight $k$. Then we plug this expansion into the second order differential equation which $h(x)$ satisfies
\beq
(1-x) x h''(x) +(1+\epsilon_3-z(2+\epsilon_1+\epsilon_2))h'(x) - (1+\epsilon_1)\epsilon_2 h(x)=0\,,
\eeq
and use the basic property of multiple polylogarithms
\begin{equation}
\frac{d}{dx} L_{0w}(x) = \frac{1}{x} L_w(x)\,,~~~\frac{d}{dx} L_{1w}(x) = \frac{1}{x-1} L_w(x)\,,
\end{equation}
to replace derivatives by action on the letters. This allows to determine
\begin{equation}
h(x) = 1 - \epsilon_2 L_1(x) + \left(\epsilon_2(\epsilon_1+\epsilon_2-\epsilon_3) L_{11}(x)-\epsilon_2(\epsilon_1-\epsilon_3) L_{01}(x)  \right) + \cdots\,.
\end{equation}
From the equation it turns out $h^{(k+1)}(x)$ can be determined in terms of $h^{(k)}(x)$, for $k=1,2,\cdots$ by the following operation
\begin{equation}
h^{(k+1)}(x) = \left(\left(-\epsilon_3 \sigma_{00}+(\epsilon_3-\epsilon_2) \sigma_{10}\right) \nabla_0+ \left((\epsilon_1-\epsilon_3) \sigma_{01}+(\epsilon_3-\epsilon_1-\epsilon_2) \sigma_{11}\right) \nabla_1 \right)h^{(k)}(x) \,,
\end{equation}
where the operators $ \nabla_0, \nabla_1$ act like a `derivative' with respect to the first letter, so that
\begin{equation}
\nabla_0 L_{0w}(x) =\nabla_1 L_{1w}(x) = L_w(x)\,,~~\nabla_1 L_{0w}(x)=\nabla_0 L_{1w}(x)=0\,,
\end{equation}
while $\sigma_{mn}$, for $m,n=0,1$ act by concatenating the letters $mn$ from the left
\begin{equation}
\sigma_{mn} L_{w}(x) =L_{mnw}(x)\,.
\end{equation}
With this, $h_1(x),h_2(x)$ can be computed to any desired order. Plugging this expansions into ${\cal J}(x,\bar x) $ it can be explicitly checked, order by order, that the resulting expression contains only single-valued multiple polylogarithms
\begin{equation}
{\cal J}(x,\bar x) = -\frac{a+b}{a b} -\frac{c}{a} {\cal L}_0(x)- \frac{c}{b}  {\cal L}_1(x)+\cdots\,.
\end{equation}

\bibliographystyle{JHEP}
\bibliography{sv_ads_veneziano}
\end{document}